%% file: commics-paper.tex
\documentclass[12pt,reqno,a4paper]{amsart}
\usepackage[utf8]{inputenc}
\usepackage[T1]{fontenc}
\usepackage[english]{babel}
\usepackage{fullpage}
\usepackage{enumitem}
\usepackage{amsmath,amssymb,amsthm,color,mathscinet}
\usepackage{framed} 
\usepackage{nameref}
\usepackage{listings}
\usepackage{siunitx}
\usepackage{etoolbox} 
\usepackage{tikz,tikz-3dplot}
\usepackage{pgfplots}
\pgfplotsset{compat=1.13}
\usepackage{pgfplotstable}
\usepgfplotslibrary{colorbrewer}
\usepackage{hyperref} 
\usepackage{chemmacros}
\usepackage{caption}
\usepackage{url}
\usepackage{xcolor}
\definecolor{pyCommentColor}{rgb}{0.204,0.396,0.643}
\input{pythonListingStyle.tex}

\def\0{\boldsymbol{0}}
\renewcommand\aa{\boldsymbol{a}}
\newcommand\ee{\boldsymbol{e}}
\def\abs#1{\vert#1\vert}
\newcommand\dx{\mathrm{d}\xx}
\newcommand\Hc{\boldsymbol{H}_{\mathrm{c}}}
\newcommand\Heff{\boldsymbol{H}_{\mathrm{eff}}}
\newcommand\Hext{\boldsymbol{H}_{\mathrm{ext}}}
\newcommand\Hs{\boldsymbol{H}_{\mathrm{s}}}
\newcommand\JJe{\boldsymbol{J}_{\mathrm{e}}}
\newcommand\Kh{\boldsymbol{\mathcal{K}}_h}
\newcommand\LL{\boldsymbol{L}}
\newcommand\meshsize{h}
\newcommand\MM{\boldsymbol{M}}
\newcommand\mm{\boldsymbol{m}}
\newcommand\nn{\boldsymbol{n}}
\newcommand\Ms{M_{\mathrm{s}}}
\newcommand\pp{\boldsymbol{p}}
\newcommand\ppsi{\boldsymbol{\psi}}
\def\prod#1#2{{\langle #1,#2\rangle}}
\def\prodh#1#2{{\langle #1,#2\rangle}_h}
\newcommand\R{\mathbb{R}}
\newcommand\timestepsize{\Delta t}
\newcommand\Trian{\mathcal{T}_h}
\newcommand\TT{\boldsymbol{T}}
\newcommand\uu{\boldsymbol{u}}
\newcommand\vv{\boldsymbol{v}}
\def\vvarphi{\boldsymbol{\varphi}}
\newcommand\xx{\boldsymbol{x}}
\newcommand\zz{\boldsymbol{z}}
\newcommand\ourmodule{\textsl{Commics}}
\newcommand\ngsolveFull{\textsl{Netgen/NGSolve}}
\newcommand\ngsolve{\textsl{NGS}}
\newcommand\bempp{\textsl{BEM++}}
\newcommand\oommf{\textsl{OOMMF}}
\title{Computational micromagnetics with Commics}
\date{\today}
\usepackage{fancyhdr}
\advance\footskip0.4cm
\advance\textheight-0.4cm
\calclayout
\makeatletter
\def\@seccntformat#1{%
  \protect\textup{\protect\@secnumfont
    \ifnum\pdfstrcmp{subsection}{#1}=0 \bfseries\fi
    \csname the#1\endcsname
    \protect\@secnumpunct
  }%
}
\makeatother
\newcommand*\patchAmsMathEnvironmentForLineno[1]{%
  \expandafter\let\csname old#1\expandafter\endcsname\csname #1\endcsname
  \expandafter\let\csname oldend#1\expandafter\endcsname\csname end#1\endcsname
  \renewenvironment{#1}%
     {\linenomath\csname old#1\endcsname}%
     {\csname oldend#1\endcsname\endlinenomath}}%
\newcommand*\patchBothAmsMathEnvironmentsForLineno[1]{%
  \patchAmsMathEnvironmentForLineno{#1}%
  \patchAmsMathEnvironmentForLineno{#1*}}%
\AtBeginDocument{%
\patchBothAmsMathEnvironmentsForLineno{equation}%
\patchBothAmsMathEnvironmentsForLineno{align}%
\patchBothAmsMathEnvironmentsForLineno{flalign}%
\patchBothAmsMathEnvironmentsForLineno{alignat}%
\patchBothAmsMathEnvironmentsForLineno{gather}%
\patchBothAmsMathEnvironmentsForLineno{multline}%
}
\usepackage[mathlines]{lineno}

\author{Carl-Martin~Pfeiler}
\author{Michele~Ruggeri}
\author{Bernhard~Stiftner}
\author{Lukas~Exl}
\author{Matthias~Hochsteger}
\author{Gino~Hrkac}
\author{Joachim~Sch\"oberl}
\author{Norbert~J.\ Mauser}
\author{Dirk~Praetorius}
\address{College of Engineering, Mathematics and Physical Sciences,
University of Exeter,
North Park Road,
Exeter, UK}
\email{g.hrkac@exeter.ac.uk}
\address{Faculty of Mathematics,
University of Vienna,
Oskar-Morgenstern-Platz 1,
1090 Vienna, Austria}
\email{michele.ruggeri@univie.ac.at}
\address{Institute for Analysis and Scientific Computing,
TU Wien,
Wiedner Hauptstrasse 8--10,
1040 Vienna, Austria}
\email{carl-martin.pfeiler@asc.tuwien.ac.at}
\email{bernhard.stiftner@asc.tuwien.ac.at}
\email{matthias.hochsteger@tuwien.ac.at}
\email{joachim.schoeberl@tuwien.ac.at}
\email{dirk.praetorius@asc.tuwien.ac.at}
\address{Wolfgang Pauli Institute c/o Faculty of Mathematics,
University of Vienna,
Oskar-Morgenstern-Platz 1,
1090 Vienna, Austria}
\email{lukas.exl@univie.ac.at}
\email{norbert.mauser@univie.ac.at}
\thanks{\emph{Acknowledgements.}
This research has been supported
by the Vienna Science and Technology Fund (WWTF)
through the project \emph{Thermally controlled magnetization dynamics} (grant MA14-44)
and by the Austrian Science Fund (FWF) through
the doctoral school \emph{Dissipation and dispersion in nonlinear PDEs} (grant W1245),
the special research program \emph{Taming complexity in partial differential systems} (grant F65),
and the project \emph{Reduced order approaches for micromagnetics} (grant P31140).
Partial support from the Engineering and Physical Sciences Research Council (EPSRC)
through the projects \emph{Picosecond dynamics of magnetic exchange springs} (grant EP/P02047X/1) and 
\emph{Coherent spin waves for emerging nanoscale magnonic logic architectures} (grant EP/L019876/1)
is thankfully acknowledged.
}
\begin{document}
\begin{abstract}
We present our open-source Python module \ourmodule{} for the study of the magnetization dynamics in ferromagnetic materials via micromagnetic simulations.
It implements state-of-the-art unconditionally convergent finite element methods for the numerical integration of the Landau--Lifshitz--Gilbert equation.
The implementation is based on the multiphysics finite element software \ngsolveFull{}.
The simulation scripts are written in Python, which leads to very readable code and direct access to extensive post-processing.
Together with documentation and example scripts, the code is freely available on GitLab.
\end{abstract}
\maketitle
\section{Introduction}\label{section:introduction}
Micromagnetism is a continuum theory for ferromagnetic materials located between Maxwell's electromagnetism and quantum theory~\cite{Brown63,Aharoni00,Kronmueller07}.
The magnetization distribution is modeled as a continuous vector field, where nonlocal magnetostatic interactions and local contributions are taken into account.
Typical micromagnetic models exhibit length scales ranging from nanometers to few micrometers, which is often infeasible for atomistic spin dynamics simulations.
In recent decades, micromagnetics evolved as a computational field, that nowadays represents a successful tool for numerical studies in materials science with important contemporary applications, e.g., data storage structures like hard disk drives~\cite{SVA+15,KOSS16}, random access memories~\cite{MSOS12}, or nanowires~\cite{SZZS00,Hertel01}, magnetologic devices~\cite{BSH+08}, soft magnetic sensor systems~\cite{BSP+17}, and high performance permanent magnets~\cite{SON+13,BOS+14,FKO+17}.
\subsection{Existing software}
In recent years, advances in computer architecture, programming environments, and numerical methods led to the development of several micromagnetic codes, which aim at the numerical integration of the fundamental equation in micromagnetics, the Landau--Lifshitz--Gilbert equation (LLG), a well-accepted model for the magnetization dynamics~\cite{LL08,Gilbert55}.
Well-known simulation packages based on finite difference discretizations~\cite{MD07} on Cartesian grids are \oommf{}~\cite{DP99}, recently extended for GPU usage~\cite{FCH+16} and endowed with a user-friendly Python interface~\cite{BPF17}, \textsl{MuMax}$^3$ (GPU)~\cite{VLD+14}, \textsl{MicroMagnum}~\cite{micromagnum} (CPU and GPU), \textsl{magnum.fd}~\cite{magnumfd}, and \textsl{Fidimag}~\cite{fidimag}.
These implementations are both memory and computationally efficient owing to the uniform mesh and the particularly advantageous utilization of the fast Fourier transform for the long range field part \cite{BRH93,AES+13}.
However, approaches based on the Finite Element Method (FEM) are geometrically more flexible~\cite{SHB+07} and provided in the scientific codes \textsl{MagPar}~\cite{SFS+03}, \textsl{TetraMag}~\cite{KWH10} and its successor \textsl{tetmag2}, \textsl{Nmag}~\cite{FFBF07} and its successor \textsl{Finmag}~\cite{finmag}, as well as in commercial software like \textsl{FastMag}~\cite{CLL+11}, \textsl{FEMME}~\cite{femme}, and \textsl{magnum.fe}~\cite{AEB+13}.
\subsection{Numerical analysis}
For an introduction to the mathematical analysis of numerical integrators for dynamic micromagnetic simulations, we refer to the monographs~\cite{Prohl01,BBN+14}, the review articles~\cite{KP06,garciaCervera2007,Cimrak08}, and the references therein.
The ultimate goal is the development of unconditionally convergent integrators, i.e., numerical schemes for which (a subsequence of) the output converges towards a weak solution of LLG in the sense of~\cite{AS92} without requiring any restrictive CFL-type coupling condition on the temporal and spatial discretization parameters.
In our work, we consider two types of methods characterized by such good theoretical properties: 
the \emph{tangent plane scheme}~\cite{Alouges08,AKS+14} and the \emph{midpoint scheme}~\cite{BP06}; see Section~\ref{section:algorithms} for more details.
\subsection{Contributions}
This work presents our novel open-source Python module \ourmodule{} (COmputational MicroMagnetICS) to perform computational studies of the magnetization dynamics in ferromagnetic materials via micromagnetic simulations.
The software is based on the multiphysics finite element software \ngsolveFull{}~\cite{ngsolve} and is made user-friendly by a high-level Python interface.
While many existing codes popular in the physics community are fairly performance optimized, they often lack a thorough mathematical convergence analysis.
In contrast to that, our implementation arises from recent results in the numerical analysis of unconditionally convergent LLG integrators.
The code is freely available on GitLab~\cite{commics} together with documentation and example scripts.
\subsection{Outline}
This work is organized as follows:
We fix the notation and the precise micromagnetic setting in Section~\ref{section:micromagnetic_setting}.
In Section~\ref{section:algorithms}, the implemented algorithms are briefly presented.
Section~\ref{section:implementation} demonstrates the exemplary use of \ngsolveFull{}~(\ngsolve{}) and discusses the integration of the boundary element library \bempp{}~\cite{SBA+15}.
Finally, Section~\ref{section:numerics} provides Python scripts for several benchmark problems, in order to verify our module and to demonstrate its usage.
\section{Micromagnetic setting}\label{section:micromagnetic_setting}
Let $\Omega \subset \R^3$ denote the volume occupied by a ferromagnet.
In micromagnetics, the quantity of interest is the magnetization $\MM: \, \Omega \to \mathbb{R}^3$ (in \si{\ampere\per\meter}).
If the temperature is constant and far below the so-called Curie temperature of the material, the modulus of the magnetization is constant, i.e., it holds that $\vert \MM \vert = \Ms$ with $\Ms>0$ being the saturation magnetization (in \si{\ampere\per\meter}).
Let $\mm := \MM / \Ms$ denote the normalized magnetization.
The magnetic state of $\Omega$ is described in terms of the magnetic Gibbs free energy (in \si{\joule})
\begin{equation} \label{eq:gibbs_energy}
\begin{split}
\mathcal{E}(\mm)
& = A  \int_{\Omega} \vert \nabla \mm \vert^2 \, \dx 
\, - \, K \int_{\Omega} (\aa \cdot \mm)^2 \, \dx
\, + \, D \int_{\Omega} (\nabla \times \mm) \cdot \mm \, \dx \\
& \quad - \, \frac{\mu_0 \Ms}{2} \int_{\Omega} \Hs \cdot \mm \, \dx
\, - \, \mu_0 \Ms \int_{\Omega} \Hext \cdot \mm \, \dx.
\end{split}
\end{equation}
The energy in~\eqref{eq:gibbs_energy} is the sum of exchange energy, uniaxial anisotropy, bulk Dzyaloshinskii--Moriya interaction (DMI), magnetostatic energy, and Zeeman contribution, respectively.
The involved material parameters and physical constants are the exchange stiffness constant $A>0$ (in \si{\joule\per\meter}), the anisotropy constant $K \geq 0$ (in \si{\joule\per\meter\cubed}), the easy axis $\aa \in \R^3$ with $\abs{\aa} = 1$ (dimensionless), the DMI constant $D \in \R$ (in \si{\joule\per\meter\squared}), and the vacuum permeability $\mu_0 =$ \SI{4 \pi e-7}{\newton\per\square\ampere}.
Moreover, $\Hext$ and $\Hs$ denote the applied external field (assumed to be unaffected by variations of $\mm$) and the stray field, respectively (both in \si{\ampere\per\meter}).
The stray field (sometimes also referred to as demagnetizing or dipolar field) solves the magnetostatic Maxwell equations
\begin{subequations}\label{eq:magnetostatic}
\begin{alignat}{2}
\nabla \cdot (\Hs + \Ms\mm\chi_{\Omega}) &= 0 &\quad& \text{in } \R^3,
\\
\nabla \times \Hs &= \0 &\quad& \text{in } \R^3,
\end{alignat}
\end{subequations}
where $(\mm\chi_{\Omega})(\xx) = \mm(\xx)$ in $\Omega$ and $(\mm\chi_{\Omega})(\xx) = 0$ elsewhere.
Stable magnetization configurations are those which minimize the magnetic Gibbs free energy~\eqref{eq:gibbs_energy}.
The dynamics towards equilibrium of the magnetization is governed by LLG
\begin{subequations} \label{eq:LLG_physical}
\begin{align}
\partial_t \mm &= - \gamma_0 \, \mm \times  \big[ \Heff(\mm) + \TT(\mm) \big]
+ \alpha \, \mm \times \partial_t \mm && \text{in } (0,\infty) \times\Omega, \label{eq:LLG_physical1}\\
\partial_{\nn}\mm &= - \frac{D}{2A} \, \mm \times \nn  && \text{on } (0,\infty) \times \partial\Omega, \label{eq:LLG_physical2} \\
\mm(0) & = \mm^0  \quad \text{ with } \abs{\mm^0} = 1 && \text{in } \Omega.  \label{eq:LLG_physical3}
\end{align}
\end{subequations}
In~\eqref{eq:LLG_physical}, $\gamma_0 =$ \SI{2.212e5}{\meter\per\ampere\per\second} is the gyromagnetic ratio of the electron, $\alpha \in (0,1]$ is the dimensionless Gilbert damping parameter, and $\nn: \partial\Omega \to \R^3$ with $\vert \nn \vert = 1$ denotes the outward-pointing unit normal vector to $\partial\Omega$.
The effective field $\Heff(\mm)$ is related to the functional derivative of the energy with respect to the magnetization and takes the form
\begin{equation*}
\begin{split}
\Heff(\mm)
:&=
- \frac{1}{\mu_0 \Ms} \frac{\delta \mathcal{E}(\mm)}{\delta \mm} \\
&= \frac{2 A}{\mu_0 \Ms} \, \Delta \mm
+ \frac{2 K }{\mu_0 \Ms} \, (\aa \cdot \mm ) \aa
- \frac{2 D}{\mu_0 \Ms} \, \nabla \times \mm 
+ \Hs
+ \Hext.
\end{split}
\end{equation*}
Finally, the term $\TT(\mm)$ collects all nonenergetic torque terms, which arise, e.g., when an electric current flows in a conducting ferromagnet.
For instance, the Oersted field $\TT(\mm) = \Hc$ (in \si{\ampere\per\meter}) is described by the magnetostatic Maxwell equations
\begin{subequations}\label{eq:magnetostatic2}
\begin{alignat}{2}
\nabla \cdot \Hc &= 0 &\quad& \text{in } \R^3,
\\
\nabla \times \Hc &=\JJe \chi_{\Omega} && \text{in } \R^3,
\end{alignat}
\end{subequations}
where $\boldsymbol{J}_{\mathrm{e}}$ denotes the electric current density (in \si{\ampere\per\meter\squared}).
Two other prominent examples are related to the so-called spin transfer torque~\cite{Slonczewski96,Berger96}, which arises in the presence of spin-polarized currents.
The Slonczewski contribution~\cite{Slonczewski96}, which takes the form
\begin{equation} \label{eq:slonczewski}
\begin{split}
& \TT(\mm) = \frac{\hbar \vert\JJe \vert G(\mm \cdot \pp,P)}{e \mu_0 \Ms d} \mm \times \pp,\\
& \qquad \text{with}
\quad
G(\mm \cdot \pp,P) = \left[ \frac{(1+P)^3 (3 + \mm \cdot \pp)}{4 P^{3/2}} - 4 \right]^{-1},
\end{split}
\end{equation}
is used for the simulation of switching processes in structures with current-perpendicular-to-plane injection geometries, e.g., magnetic multilayers.
The involved physical quantities are the reduced Planck constant $\hbar>0$ (in \si{\joule\second}), the elementary charge $e>0$ (in~\si{\ampere\second}), a dimensionless polarization parameter $P \in (0,1)$, a dimensionless unit vector $\pp \in \R^3$ representing the magnetization of a uniformly-magnetized polarizing layer (the so-called fixed layer), and the thickness $d>0$ of the so-called free layer (in \si{\meter}).
The Zhang--Li contribution~\cite{ZL04,TNM+05} is used for the simulation of the current-driven motion of domain walls in single-phase samples characterized by current-in-plane injection geometries and, according to~\cite{ZL04} (resp., \cite{TNM+05}), takes the form
\begin{subequations} \label{eq:zhang-li}
\begin{align}
\label{eq:zhang-li1}
& \TT(\mm) = - \frac{1}{\gamma_0} [\mm \times (\uu \cdot \nabla) \mm + \xi (\uu \cdot \nabla)\mm ], \\
& \qquad \text{with}
\quad
\uu = - \frac{P \mu_B}{e \Ms (1 + \xi^2)}\JJe
\quad
\left( \text{resp., } \uu = - \frac{P g_\mathrm{e} \mu_B}{2 e \Ms}\JJe \right).
\end{align}
\end{subequations}
The involved physical quantities are the spin velocity vector $\uu \in \R^3$ (in \si{\meter\per\second}), the dimensionless ratio of nonadiabaticity $\xi >0$, the Bohr magneton $\mu_B>0$ (in \si{\ampere\meter\squared}), and the dimensionless g-factor of the electron $g_\mathrm{e} \approx 2$.

In~\eqref{eq:gibbs_energy} and~\eqref{eq:LLG_physical}, to fix the ideas, we considered the case of the bulk DMI as a prototype for chiral interactions~\cite{Dzyaloshinskii58, Moriya60}.
However, our implementation also covers the case of the interfacial DMI; see, e.g.,~\cite{CL98,scrtf2013}.
If $D=0$ (no chiral interaction), then the boundary conditions~\eqref{eq:LLG_physical2} become homogeneous Neumann boundary conditions.
\section{Algorithms} \label{section:algorithms}
The algorithms implemented in our Python module \ourmodule{} employ a uniform partition of the time interval with constant time-step size $\timestepsize>0$.
For the spatial discretization, we consider a tetrahedral mesh $\Trian$ of the ferromagnet $\Omega$ with mesh size $\meshsize>0$.
The associated FEM space of piecewise affine and globally continuous functions reads
\begin{equation}
\label{eq:fespace_1D}
V_h := \{ \varphi_h\colon\Omega\to\R \text{ continuous} : 
\varphi_h\vert_K \textrm{ is affine for all elements } K \in  \Trian \}.
\end{equation}
For each time-step $n = 0,1,2,\dots$, we seek for approximations
\begin{equation*}
(V_h)^3 \ni \mm_h^n \approx \mm(n\timestepsize) \quad \text{such that} \quad \abs{\mm_h^n(\zz)} = 1 \text{ for all nodes } \zz \text{ of } \Trian,
\end{equation*}
i.e., for any time-step, the approximate magnetization satisfies the unit-length constraint $\abs{\mm}=1$ at the nodes of the mesh.
\subsection{Tangent plane scheme}
Tangent plane schemes (sometimes also referred to as \emph{projection methods}) are based on variational formulations of the equivalent form of LLG
\begin{equation*}
\alpha \, \partial_t \mm
+ \mm \times \partial_t \mm
= \gamma_0 ( \Heff(\mm) + \TT(\mm) )
-  \gamma_0 [ (\Heff(\mm) + \TT(\mm)) \cdot \mm ] \mm.
\end{equation*}
The orthogonality $\mm \cdot \partial_t \mm = 0$, which characterizes any solution of~\eqref{eq:LLG_physical1}, is enforced at the discrete level by considering the discrete tangent space
\begin{equation*}
\Kh(\mm_h^n)
:= 
\{ \vvarphi_h \in (V_h)^3 : \vvarphi_h(\zz) \cdot \mm_h^n(\zz) = 0 \text{ for all nodes } \zz \text{ of } \Trian \} \subset (V_h)^3.
\end{equation*}
For each time-step, one has to solve a constrained linear system to compute $\vv_h^n \approx \partial_t \mm(n\timestepsize)$ in $\Kh(\mm_h^n)$.
With $\mm_h^n$ and $\vv_h^n$ at hand, one then computes
\begin{equation*}
\mm_h^{n+1}\in (V_h)^3 \quad \textrm{by} \quad
\mm_h^{n+1}(\zz) := 
\frac{\mm_h^{n}(\zz) + \timestepsize \, \vv_h^{n}(\zz)}{\abs{\mm_h^{n}(\zz) + \timestepsize \, \vv_h^{n}(\zz)}} \quad
\text{for all nodes $\zz$ of } \Trian.
\end{equation*}
The original tangent plane scheme from~\cite{Alouges08} is formally first-order in time and was analyzed for the energy being only the exchange contribution.
The scheme was extended to general lower-order effective field contributions~\cite{AKT12,BSF+14}, DMI~\cite{hpprss2017}, and the coupling with other partial differential equations, e.g., various forms of Maxwell's equations~\cite{LT13,BPP15,LPP+15,FT17}, spin diffusion~\cite{AHP+14}, and magnetostriction \cite{BPP+14}.
A projection-free version of the method was analyzed in~\cite{AHP+14, Ruggeri16}.
In a variant from~\cite{AKS+14,DPP+17}, the formal convergence order in time has been increased from one to two.
Effective solution strategies and preconditioning for the resulting constrained linear system have recently been proposed in~\cite{kpprs2018}.
\subsection{Midpoint scheme} \label{sec:midpointScheme}
The midpoint scheme is based on a variational formulation of the Gilbert form~\eqref{eq:LLG_physical1} of LLG.
It consists of two fundamental ingredients:
the implicit midpoint rule in time and the mass-lumped $L^2$-product in space, defined as
\begin{equation} \label{eq:massLumping}
\prodh{\vvarphi}{\ppsi} := \int_{\Omega} \mathcal{I}_h ( \vvarphi \cdot \ppsi ) \dx
\, \approx  \,
\prod{\vvarphi}{\ppsi}_{\LL^2(\Omega)} \quad \textrm{for all } \vvarphi, \ppsi\colon\Omega\to\R^3\textrm{ continuous}.
\end{equation}
Here, $\mathcal{I}_h$ is the standard nodal interpolant associated with $\Trian$.
The resulting scheme is second-order in time, inherently preserves the unit-length constraint and the energy of the solutions, but requires the solution of one \emph{nonlinear} system for $\mm_h^{n+1/2} := (\mm_h^{n+1} + \mm_h^n)/2 \in (V_h)^3$ per time-step.
The midpoint scheme was proposed and analyzed in~\cite{BP06}.
The scheme was extended to lower-order terms~\cite{PRS18}, the coupling with the Maxwell equation~\cite{BBP08}, and a variant of LLG in heat-assisted magnetic recording~\cite{BPS09,BPS12}.
The resulting nonlinear system is usually solved with constraint-preserving fixed-point iterations, which, however, spoil the unconditional convergence.
\subsection{Magnetostatic Maxwell equations} \label{subsection:fk}
For the computation of stray field and Oersted field, i.e., for the numerical solution of the magnetostatic Maxwell equations~\eqref{eq:magnetostatic} and~\eqref{eq:magnetostatic2}, we follow a hybrid FEM-BEM approach, which combines FEM with the boundary element method (BEM); see~\cite{FK90,hk2014}. The method uses the superposition principle and computes the magnetic scalar potential $u$ such that $\Hs = - \nabla u$ and the Oersted field $\Hc$ by splitting the problem into two parts, where BEM techniques for the evaluation of the double-layer potential are employed; see, e.g., \cite[Chapter~4.1]{PRS18} for details.
\section{Implementation}\label{section:implementation}
Our Python module \ourmodule{} is based on the \ngsolveFull{}~\cite{ngsolve}~(\ngsolve{}) FEM software and provides a tool to perform micromagnetic simulations with the algorithms described
in Section~\ref{section:algorithms} for a variety of energy contributions and dissipative effects.
It is purely Python-based, provides extensive simulation data for reproducibility and post-processing, and automatically takes care of, e.g., the definition and the assembly of underlying (bi)linear
forms as well as the numerical solution of the arising linear and nonlinear systems.
Although not needed for the use of \ourmodule{}, this section advertises some of the core features of \ngsolve{}, as well as the coupling of \ngsolve{} with \bempp{}.
\subsection{Basic features}\label{sec:basicFeatures}
Geometry handling, mesh-generation, FEM spaces, and assembly routines are intentionally hidden from the user of \ourmodule{} and are internally covered in the \ngsolve{} framework.
For instance, given an \ngsolve{} object \texttt{mesh}, representing a tetrahedral mesh of the domain $\Omega$, the discrete vector-valued product space $(V_h)^3$ is simply generated by
\begin{lstlisting}
Vh!3! = VectorH!1!(mesh, order=1)
\end{lstlisting}
where the syntax \texttt{VectorH1} indicates that \texttt{Vh3} is a proper subspace of the vector-valued Sobolev space $(H^1(\Omega))^3$.
\subsection{Symbolic (bi)linear forms}
Although the use of \ourmodule{} requires minimal knowledge of \ngsolve{}, we shortly describe one feature of the library, namely the definition of (bi)linear forms:
\ngsolve{} allows the symbolic definition of (time-dependent) (bi)linear forms.
For instance, given the space \texttt{Vh3} defined in Section~\ref{sec:basicFeatures}, the LLG-specific cross-product bilinear form
\begin{align}\label{eq:skew_example}
\prod{\mm_h^n \times \ppsi_h}{\vvarphi_h}_{\LL^2(\Omega)} \quad \textrm{for all } \vvarphi_h, \ppsi_h \in (V_h)^3
\end{align}
is symbolically defined on the Python level by the following code snippet:
\begin{lstlisting}
# grid-, trial- and testfunction
psi = Vh!3!.TrialFunction()
phi = Vh!3!.TestFunction()
m = GridFunction(Vh!3!)
# bilinear form
LHS = BilinearForm(Vh!3!)
ir = IntegrationRule(TET, order=3)
LHS += SymbolicBFI(Cross(m, psi) * phi, intrule=ir)
\end{lstlisting}
In the last line, \texttt{SymbolicBFI} with the test function \texttt{phi} and the trial function \texttt{psi} realizes the bilinear form~\eqref{eq:skew_example} and adds it to the left-hand side \texttt{LHS} for further handling.
Note that via the \texttt{intrule} option, the order of the quadrature is explicitly set to~$\num{3}$, since the realization of the bilinear form corresponds to the exact integration of piecewise cubic polynomials on tetrahedra, hence also the parameter \texttt{TET}.
For the midpoint scheme, the corresponding bilinear form employs the mass-lumped $L^2$-product from~\eqref{eq:massLumping}, i.e.,
\begin{equation}\label{eq:skew_example_massLumped}
\prod{\mm_h^n \times \ppsi_h}{\vvarphi_h}_{h} \quad \textrm{for all } \vvarphi_h, \ppsi_h \in (V_h)^3.
\end{equation}
This bilinear form can be defined in the same way as above specifying the corresponding quadrature rule on the reference tetrahedron $\operatorname{conv}\{\0, \ee_1, \ee_2, \ee_3\}$:
\begin{lstlisting}
ir = IntegrationRule(points=[(0,0,0), (1,0,0), (0,1,0), (0,0,1)], \
                     weights=[1/24, 1/24, 1/24, 1/24])
\end{lstlisting}
\subsection{Coupling with \bempp{}}
For the approximate computation of the magnetostatic fields with the hybrid FEM-BEM approach from Section~\ref{subsection:fk}, the evaluation of the double-layer potential is required.
To that end, we incorporate into \ourmodule{} the corresponding functionality of the BEM software \bempp{}~\cite{SBA+15}, including matrix compression techniques~\cite{Hackbusch1999}.
The coupling of \ngsolve{} and \bempp{} is done on the Python level with the \textsl{ngbem} module~\cite{Rieder}; see Appendix~\ref{appendix:fk}.
\section{Using Commics: Standard problems and numerical experiments}\label{section:numerics}
In this section, we present some numerical experiments performed with \ourmodule{}.
For each proposed example, we also include the executable Python script to run the simulation.
\subsection{Using \ourmodule{}}
To run a micromagnetic simulation with \ourmodule{}, the user has to define an object of class \texttt{commics.Integrator} and call its \texttt{Integrate} method.

The essential inputs to initialize such an object are the geometry as a \texttt{commics.Geometry} object defining the ferromagnetic domain and the meshing strategy (see Section~\ref{subsection:meshing}), as well as an object of type \texttt{commics.Parameters} specifying, among other things, material parameters, an applied field or current, the initial magnetization state, and the time discretization strategy.
Moreover, the desired time integration scheme has to be chosen.
The algorithms for the numerical integration of LLG implemented in \ourmodule{} described in Section~\ref{section:algorithms} can be selected in the following way:
\begin{itemize}
\item The first-order tangent plane scheme from~\cite{Alouges08} with explicit integration of the lower-order terms in time~\cite{AKT12, BSF+14} is provided as \texttt{TPS1}.
\item The projection-free tangent plane scheme (with explicit integration of the lower-order terms in time) from~\cite{AHP+14, Ruggeri16} is provided as \texttt{TPS1PF}.
\item The second-order tangent plane scheme from~\cite{AKS+14} and its improved version from~\cite{DPP+17} are available as \texttt{TPS2} and \texttt{TPS2AB}, respectively.
\item The midpoint scheme from~\cite{BP06} and its improved version from~\cite{PRS18} are provided as \texttt{MPS} and \texttt{MPSAB}, respectively.
\end{itemize}
Various example scripts are provided subsequently in this section.
Further details can be accessed with the Python built-in \texttt{help()} function available for \ourmodule{} classes.
\subsection{Geometry specification and meshing}
\label{subsection:meshing}
\ourmodule{} provides several ways to specify a geometry and to generate a corresponding mesh:
\begin{itemize}
\item For complex geometries, the submodule \texttt{netgen.csg} of \ngsolve{} provides a rich number of possibilities to define geometries; see \cite{ngsolve}.
\item For geometries often encountered in micromagnetics, e.g., cuboids and disks, one can simply provide the dimensions of the sample, a scale factor, and the desired maximum mesh size; see, e.g., the code snippets in Section~\ref{subsection:sState} and Section~\ref{subsection:helimag}.
Then, \texttt{netgen.csg} will automatically be used with appropriate settings.
\item Existing \ngsolve{} meshes/geometries (stored as \texttt{.vol}-files) can simply be loaded; see, e.g., the code snippet in Section~\ref{subsection:switching}.
\end{itemize}
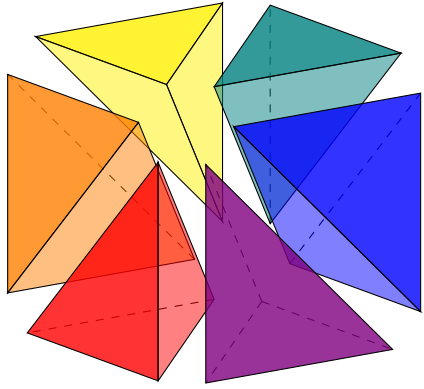
\begin{figure}[ht]
\tdplotsetmaincoords{75}{-35}
\pgfmathsetmacro{\shift}{0.15}
\centering
\input{pics/meshing/structured.tex}
\caption{Uniform decomposition of a cube into six tetrahedra.}
\label{fig:mesh}
\end{figure}
Meshes automatically generated by \ngsolve{} are unstructured and obtained by the advancing front method; see~\cite{schoeberl1997} for details.
However, due to mesh quality and shape optimizations, these meshes do not necessarily satisfy the prescribed maximum mesh size.
\ourmodule{} provides two possibilities to bypass this drawback:
For cuboidal geometries, setting the \ourmodule{} option \texttt{structuredMesh=True} allows for structured meshes.
First, the sample is uniformly split into cuboidal cells of prescribed size.
Then, each cell is split into six tetrahedra in such a way that any tetrahedron has three mutually perpendicular edges; see Figure~\ref{fig:mesh}.
This strategy is for example used in Section~\ref{subsection:mumag5}.
For general geometries, \ourmodule{} provides the means to repeatedly generate a new mesh (by prescribing a smaller and smaller mesh size each time) in \ngsolve{}, until the initial mesh size specification is satisfied.
This strategy can be enabled by setting \texttt{forceNetgenMeshSize=True} as done in Section~\ref{subsection:helimag}.

\subsection{\texorpdfstring{$\mu$}{mu}MAG standard problem~\#4}
\label{subsection:mumag4}
To describe the key aspects of a \ourmodule{} script, we consider the $\mu$MAG standard problem~\#4~\cite{mumag}.

The objective is the simulation of the magnetization dynamics in a thin permalloy film of dimensions $\SI{500}{\nm}\times\SI{125}{\nm}\times\SI{3}{\nm}$ under the influence of a constant applied external field.
We split the experiment into two parts:
In the first stage, we obtain the so-called equilibrium S-state, which is saved to serve as the initial configuration for the second stage, where the switching dynamics is simulated.
\subsubsection{Obtaining the S-state}
\label{subsection:sState}
We consider a structured tetrahedral mesh of the given cuboid into cells of size $\meshsize_x\times\meshsize_y\times\meshsize_z$, which are decomposed into tetrahedra as depicted in Figure~\ref{fig:mesh}. The dimension of the cells is chosen as $\meshsize_x=\meshsize_y=\SI{125/69}{\nm}$ and $\meshsize_z=\SI{1.5}{\nm}$. This corresponds to $\num{228528}$ elements with diameter $\meshsize=\SI{2.97}{\nm}$, $\num{58170}$ vertices, as well as $\num{78936}$ surface elements.
The material parameters of permalloy read $M_s=\SI{8e05}{\A\per\m}$, $A=\SI{1.3e-11}{\J\per\m}$, $D=\SI{0}{\J\per\m^2}$, and $K=\SI{0}{\J\per\m^3}$.
To speed up the process, we deliberately choose the large value $\alpha=1$ for the Gilbert damping parameter.
For the simulation, we use a constant time-step size $\timestepsize = \SI{0.1}{\ps}$.

The problem description suggests to obtain the S-state by applying a slowly reducing external field pointing in the $(1,1,1)$-direction.
We start with a uniform initial state $\mm_h^0\equiv(1,0,0)$ and let the magnitude $\abs{\Hext}$ of the external field decrease linearly from $\num{30}/\mu_0$ to $\SI{0}{\milli\tesla}$ over a period of $\SI{1}{\ns}$.
In \ourmodule{} scripts, non-constant fields can be described using time- and space-dependent Python \texttt{lambda}-functions.
Further, we relax the system for $\SI{1}{\ns}$ without applying any external field and store the obtained S-state as \texttt{sp4sState.vtk} for later use.
\begin{lstlisting}[]
from commics import *
# specify geometry and parameters
h_xy, h_z = 125/69, 1.5
geo = Geometry(geometry=Cuboid(500, 125, 3), meshSize=(h_xy, h_xy, h_z), \
               structuredMesh=True, scaling=1e-9)
par = Parameters(A=1.3e-11, Ms=8e+5, K=0, D=0, gamma!0!=2.211e+5, \
                 alpha=1.0, m!0!=(1.0, 0.0, 0.0), \
                 T_start=0, T_end=1e-9, timeStepSize=0.1e-12)
# define time dependent applied field via lambda function
from math import sqrt
field = lambda t,x,y,z : (par.T_end - t) / (par.T_end - par.T_start) \
        * 30.0e-3 / par.mu!0! / sqrt(3.0)
par.H_ext = (field, field, field)
# define integrator and run simulation from T_start to T_end
sp!4! = Integrator(scheme=TPS!2!AB, geometry=geo, parameters=par)
sp!4!.Integrate()
# Relax for another nanosecond and save the mesh and the result
sp!4!.Integrate(duration=1e-9, relax=True)
sp!4!.SaveMesh("sp4mesh")
sp!4!.SaveMagnetization("sp4sState")
\end{lstlisting}
\subsubsection{Switching}
\label{subsection:switching}
We assume that the folder \texttt{data} contains the two files \texttt{sp4sState.vtk} and \texttt{sp4mesh.vol} saved from the simulation described in Section~\ref{subsection:sState}.
As stated in the problem description, we choose $\alpha=0.02$ and set the external field to $\Hext = (-24.6, 4.3, 0)/\mu_0$ \si{\milli\tesla}.
Then, we run the simulation for $\SI{3}{\ns}$.
\begin{lstlisting}
from commics import *
# specify geometry and parameters
geo = Geometry(geometry="data/sp4mesh.vol", scaling=1e-9)
par = Parameters(A=1.3e-11, Ms=8e+5, K=0, D=0, gamma!0!=2.211e+05, \
                 alpha=0.02, timeStepSize=0.1e-12, m!0!="data/sp4sState.vtk")
par.H_ext = (-24.6e-3/par.mu!0!, 4.3e-3/par.mu!0!, 0.0)
# define integrator and run simulation
sp!4! = Integrator(scheme=TPS!2!AB, geometry=geo, parameters=par)
sp!4!.Integrate(duration=3e-9)
\end{lstlisting}
\begin{figure}[ht]
\centering
\input{plots/sp4/avg_combined.tex}
\caption{$\mu$MAG standard problem~\#4 from Section~\ref{subsection:mumag4}: Time evolution of the spatially averaged magnetization components computed with \ourmodule{} (\texttt{TPS2AB} and \texttt{MPS}) compared to the results of \oommf{}.}
\label{fig:mumag4_avg}
\end{figure}
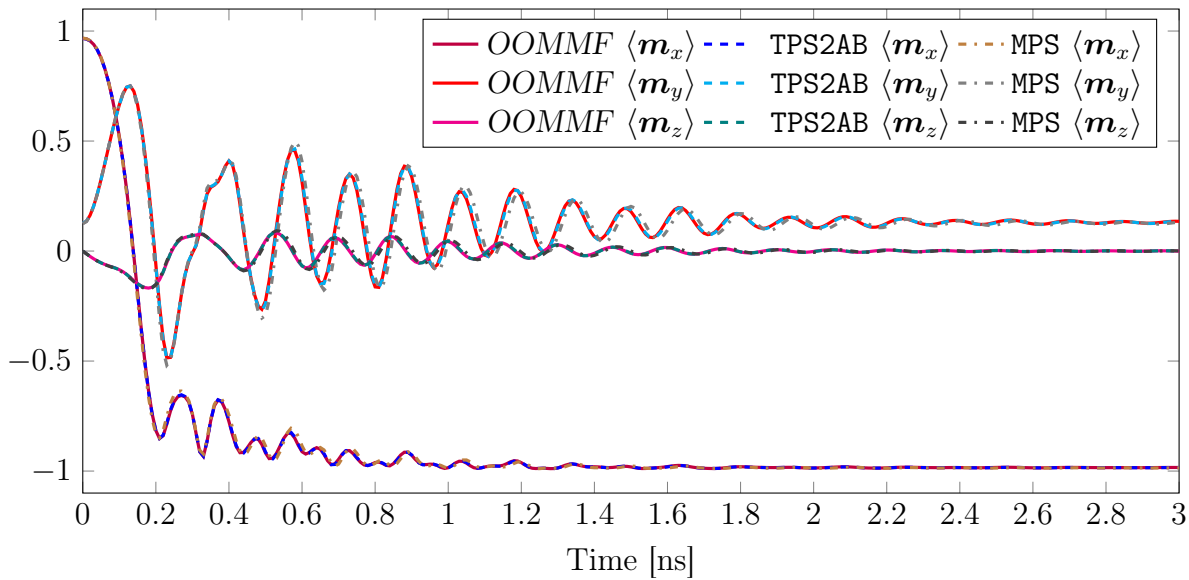
\begin{figure}[ht]
\centering
\input{pics/sp4/structured_3nm_tps2ab_xZeroColX.tex}
\caption{$\mu$MAG standard problem~\#4 from Section~\ref{subsection:mumag4}: Snapshot of the magnetization when the $x$-component of the spatially averaged magnetization first crosses zero ($t=\SI{138.2}{\ps}$).}
\label{fig:sp4_xZeroColorX}
\end{figure}
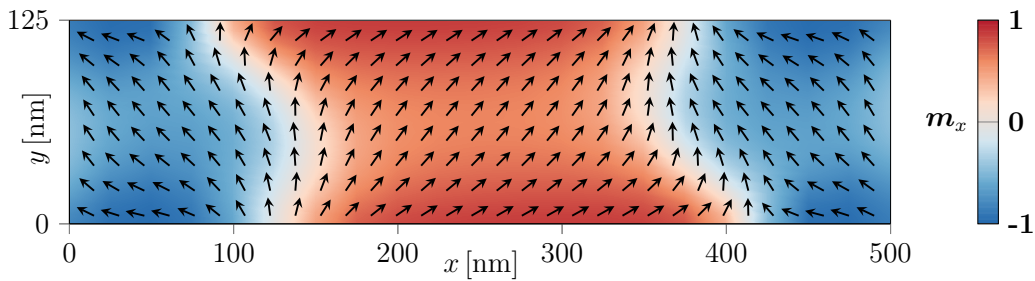
For comparison, the desired output of this benchmark problem is the evolution of the $x$-, $y$- and $z$-component of the spatially averaged magnetization.
Figure~\ref{fig:mumag4_avg} shows, that our results match those computed by the finite difference code \oommf{}~\cite{DP99} available on the $\mu$MAG homepage~\cite{mumag}.
Further, Figure~\ref{fig:sp4_xZeroColorX} visualizes the magnetization at the time when the $x$-component of the spatially averaged magnetization first crosses zero.
\subsubsection{Meshing strategy and integrator}
In Section~\ref{subsection:sState},  we considered a structured mesh and the \texttt{TPS2AB} integrator from~\cite{DPP+17}.
To compare the results, we additionally repeat the simulation on an unstructured mesh using the midpoint scheme:
To simulate the dynamics on an unstructured mesh generated by \ngsolve{}, we replace the definition of the geometry in Section~\ref{subsection:sState} by
\begin{lstlisting}
geo = Geometry(geometry=Cuboid(500, 125, 3), meshSize=3, scaling=1e-9)
\end{lstlisting}
This results in an unstructured mesh containing \num{48792} elements, \num{16682} vertices and \num{33360} surface elements, which corresponds to an actual mesh size of \SI{5.19}{\nm}.
To repeat the simulation using the midpoint scheme from~\cite{BP06}, we replace the definition of the integrator in Section~\ref{subsection:sState} and Section~\ref{subsection:switching} by
\begin{lstlisting}
sp!4! = Integrator(scheme=MPS, geometry=geo, parameters=par)
\end{lstlisting}
Although the mesh size is close to the exchange length of the material (\SI{5.69}{\nm}), qualitatively the results match those computed by \oommf{} well; see Figure~\ref{fig:mumag4_avg}.
\subsection{\texorpdfstring{$\mu$}{mu}MAG standard problem~\#5}
\label{subsection:mumag5}
The spintronic extensions of LLG from~\cite{ZL04,TNM+05} are the subject of the $\mu$MAG standard problem~\#5~\cite{mumag}.
The sample under consideration is a permalloy film with dimensions $\SI{100}{\nm}\times\SI{100}{\nm}\times\SI{10}{\nm}$ aligned with the $x$, $y$, and $z$ axes of a Cartesian coordinate system, with origin at the center of the film.
We consider the same material parameters as in Section~\ref{subsection:mumag4} and $\alpha=0.1$.
The initial state is obtained by solving~\eqref{eq:LLG_physical} with $\TT \equiv \0$ and $\mm^0(x,y,z) = (-y,x,R) / \sqrt{x^2 + y^2 + R^2}$ with $R=\SI{10}{nm}$ and maximal damping $\alpha=1$ for $\SI{1}{\ns}$, which is a sufficiently long time for the system to reach equilibrium.
Given $P\JJe = (\num{1e12}, \num{0}, \num{0})~\si{\ampere\per\meter\squared}$ and $\xi=$ \num{0.05}, we set $\TT$ according to the expression in~\eqref{eq:zhang-li1}.
Then, we solve~\eqref{eq:LLG_physical} with the relaxed magnetization configuration as initial condition for \SI{8}{\nano\second},
which turns out to be a sufficiently long time to reach the new equilibrium; see Figure~\ref{fig:sp5_evolution_picture} and Figure~\ref{fig:sp5_evolution_plot}.
We consider a structured tetrahedral mesh of the given cuboid into cells of size $\meshsize_x\times\meshsize_y\times\meshsize_z$, which are decomposed into tetrahedra as depicted in Figure~\ref{fig:mesh}. The dimension of the cells is chosen as $\meshsize_x=\meshsize_y=\meshsize_z=\SI{5/3}{\nm}$. This corresponds to $\num{129600}$ elements with diameter $\meshsize=\SI{2.89}{\nm}$, $\num{26047}$ vertices, as well as $\num{17280}$ surface elements.
\begin{lstlisting}[]
from commics import *
# specify geometry and parameters
h = 5 / 3
geo = Geometry(geometry=Cuboid((-50,-50,-5), (50,50,5)), meshSize=(h,h,h), \
               structuredMesh=True, scaling=1.0e-09)
par = Parameters(A=1.3e-11, Ms=8.0e+05, K=0.0, gamma!0!=2.211e+05, alpha=1.0, \
                 spintronicsCoupling="zhangLi", g=2.0, P=1.0, \
                 Je=(1.0e+12, 0.0, 0.0), xi=0.05, \
                 timeStepSize=0.1e-12, T_start=-1.0e-09)
# space dependent initial condition (scaled domain); normalized automatically
from ngsolve import x, y
par.m!0! = (-y, x, 10.0)
# define integrator and relax configuration
sp!5! = Integrator(scheme=TPS!2!AB, geometry=geo, parameters=par)
sp!5!.Integrate(duration=1.0e-09, relax=True)
# set damping alpha and run simulation with specified current
sp!5!.SetParameter_alpha(0.1)
sp!5!.Integrate(duration=8.0e-09)
\end{lstlisting}
\begin{figure}[ht]
\centering
\hspace{-6cm}
\begin{minipage}{13cm}
\centering
\raisebox{-0.5\height}{\input{pics/sp5/sp5_mag_0ps.tex}}
\hspace*{3mm}
\raisebox{-0.5\height}{\input{pics/sp5/sp5_mag_500ps.tex}}
\hspace*{3mm}
\raisebox{-0.5\height}{\input{pics/sp5/sp5_mag_1000ps.tex}}
\raisebox{-0.5\height}{\input{pics/sp5/sp5_mag_1500ps.tex}}
\hspace*{3mm}
\raisebox{-0.5\height}{\input{pics/sp5/sp5_mag_2000ps.tex}}
\hspace*{3mm}
\raisebox{-0.5\height}{\input{pics/sp5/sp5_mag_8000ps.tex}}
\end{minipage}
\hspace{-7.5cm}
\begin{minipage}{2cm}
\raisebox{0.125\height}{\input{pics/sp5/sp5_colorbar.tex}}
\end{minipage}
\caption{$\mu$MAG standard problem~\#5:
Magnetization in the $xy$-plane viewed from the top at different times.
Starting from the relaxed configuration at $t=\SI{0.0}{\ns}$, the vortex (red) follows a spiral-like motion.
After $t=\SI{8.0}{\ns}$ no further movements are observed.}
\label{fig:sp5_evolution_picture}
\end{figure}
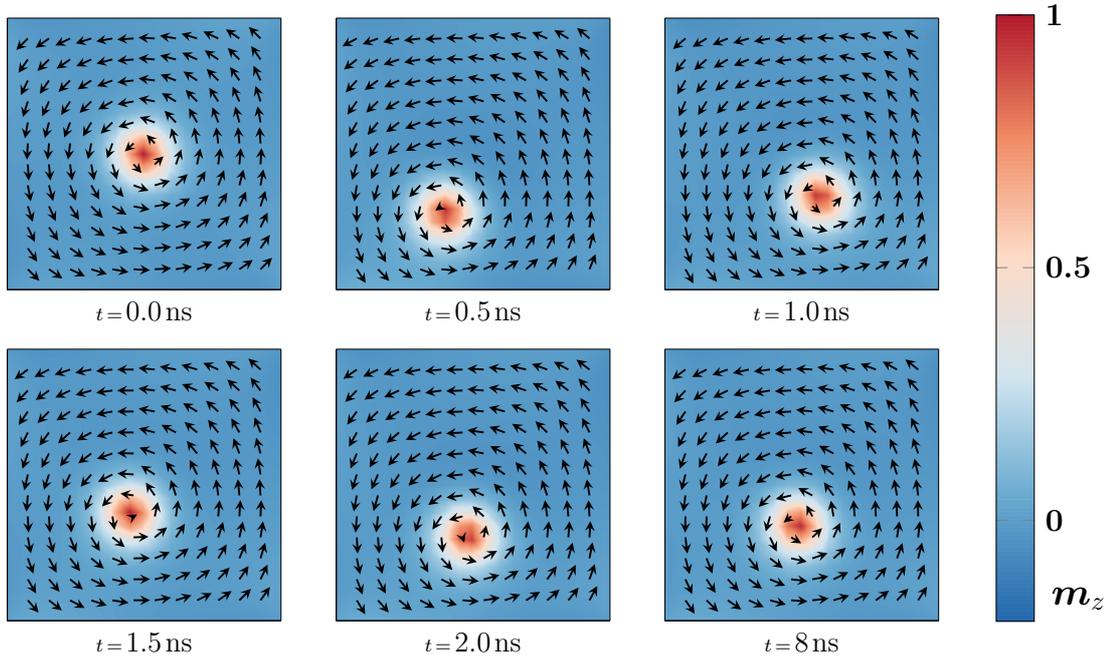
\begin{figure}[ht]
\centering
\input{plots/sp5/averages.tex}
\caption{$\mu$MAG standard problem~\#5: Evolution of the spatially averaged $x$- and $y$-component of $\mm$.}
\label{fig:sp5_evolution_plot}
\end{figure}
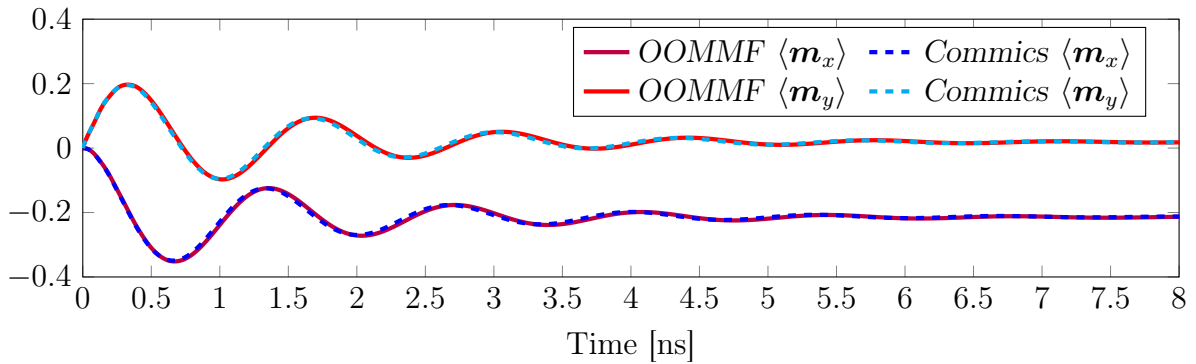
\subsection{Standard problem for ferromagnetic resonance simulations}\label{subsection:fmr}
Ferromagnetic resonance (FMR) is a well-established experimental technique for the study of ferromagnetic materials.
A typical application of FMR consists in perturbing the magnetization of a system from its equilibrium by a sufficiently weak excitation and studying the induced magnetization dynamics, which is basically made of damped oscillations around the initial equilibrium.
The resulting resonance frequencies and the eigenmodes of the system give some insights on the magnetic properties of the material and are used, e.g., for the experimental measurement of model parameters like the Gilbert damping constant or the saturation magnetization~\cite{farle1998,ms2005}.

In this section, we compute with \ourmodule{} a problem for FMR simulations recently proposed in~\cite{BBA+17}.
The computational domain is a cuboid of permalloy with dimensions $\SI{120}{\nm}\times\SI{120}{\nm}\times\SI{10}{\nm}$.
The material parameters are the same as in Section~\ref{subsection:mumag4}.
During the first stage, we set $\alpha=1$ and consider a constant applied external field of magnitude $\abs{\Hext} =$ \SI{8e4}{\ampere\per\meter} pointing in the direction $(1,0.715,0)$.
We initialize the system with a uniform ferromagnetic state $\mm_h^0\equiv(1,0,0)$ and let the system evolve for \SI{5}{\nano\second}.
The resulting state is then used as initial condition for the second stage, in which we set $\alpha=0.008$, change the direction of the applied external field to $(1,0.7,0)$ but keep $\abs{\Hext} =$ \SI{8e4}{\ampere\per\meter} , and let the system evolve to the new equilibrium for \SI{20}{\nano\second}.
We consider a structured tetrahedral mesh of the given cuboid into cells of size $\meshsize_x\times\meshsize_y\times\meshsize_z$, which are decomposed into tetrahedra as depicted in Figure~\ref{fig:mesh}. The dimension of the cells is chosen as $\meshsize_x=\meshsize_y=\meshsize_z=\SI{2}{\nm}$. This corresponds to $\num{108000}$ elements, $\num{22326}$ vertices, as well as $\num{16800}$ surface elements, and yields a mesh size of $\meshsize=\SI{3.46}{\nm}$.
We compare our results obtained with \ourmodule{} to those presented in \cite{BBA+17}.
There, the authors use the finite difference code \oommf{}~\cite{DP99} and investigate the evolution of the $y$-component of the spatially averaged magnetization, as well as its power spectrum $S_y$. The power spectrum is obtained by a discrete Fourier transform as described in~\cite[Section~2.3.1]{BBA+17}.
Our results match well with those of \cite{BBA+17}; see Figure~\ref{fig:fmr_avg_y} and Figure~\ref{fig:fmr_dft_y}.
\clearpage
\begin{lstlisting}
from commics import *
# specify geometry and parameters
geo = Geometry(geometry=Cuboid(120, 120, 10), meshSize=(2,2,2), \
               structuredMesh=True, scaling=1.0e-09)
par = Parameters(A=1.3e-11, Ms=8.0e+05, K=0.0, gamma!0!=2.210173e+05, \
                 alpha=1.0, timeStepSize=0.1e-12, T_start=-5.0e-9, \
                 m!0!=(0, 0, 1))
# define integrator, choose number of threads, specify folder for results
fmr = Integrator(scheme=TPS!2!AB, geometry=geo, parameters=par, numthreads=8)
fmr.SetResultsFolder("FMR_Result")
# obtain initial condition
amplitude = 80.0e+03
e = (1.0, 0.715, 0.0)
e_length = (sum(e[j]**2 for j in range(3)))**0.5
H_ext = (amplitude*e[0]/e_length, amplitude*e[1]/e_length, 0.0)
fmr.SetParameter_H_ext(H_ext)
fmr.Integrate(duration=5.0e-09)
# change direction of applied field and run simulation
fmr.SetParameter_alpha(0.008)
e = (1.0, 0.7, 0.0)
e_length = (sum(e[j]**2 for j in range(3)))**0.5
H_ext = (amplitude*e[0]/e_length, amplitude*e[1]/e_length, 0.0)
fmr.SetParameter_H_ext(H_ext)
fmr.Integrate(duration=20.0e-09)
\end{lstlisting}
\begin{figure}[ht]
\centering
\input{plots/fmr/avg_y.tex}
\caption{Ferromagnetic resonance simulation from Section~\ref{subsection:fmr}: Time evolution of $\langle \mm_y \rangle$ obtained with \ourmodule{} compared to the results computed with \oommf{}.}
\label{fig:fmr_avg_y}
\end{figure}
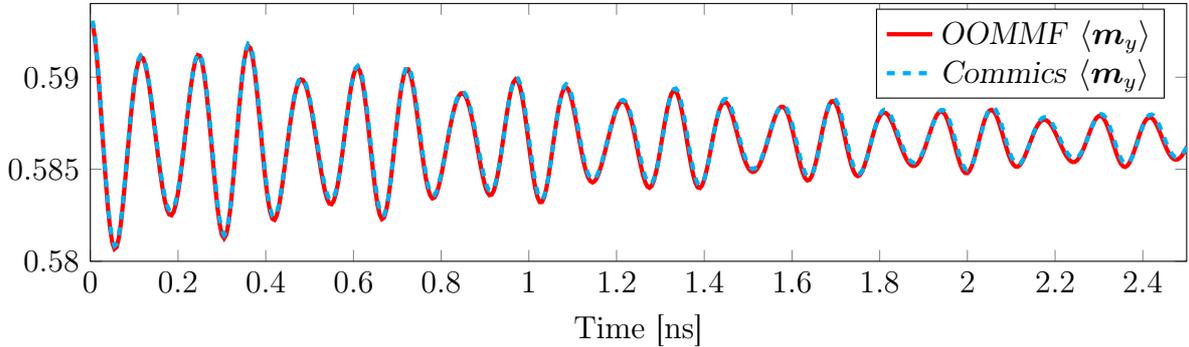
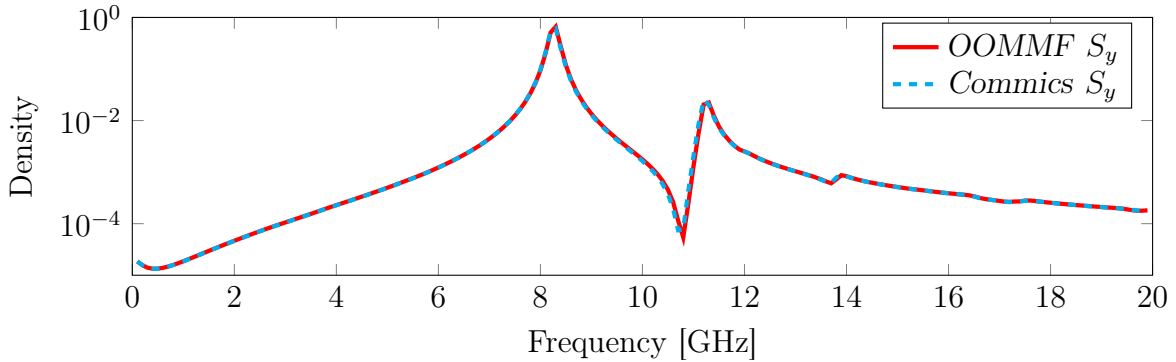
\begin{figure}[ht]
\centering
\input{plots/fmr/dft_y.tex}
\caption{Ferromagnetic resonance simulation from Section~\ref{subsection:fmr}: Power spectrum $S_y$ obtained by discrete Fourier transform of the spatially averaged $y$-component of the magnetization $\langle \mm_y \rangle$.}
\label{fig:fmr_dft_y}
\end{figure}
\subsection{Current-induced dynamics of skyrmions in nanodisks}\label{subsection:helimag}
With this experiment, we aim to show how \ourmodule{} can be used to numerically investigate the stability and the induced dynamics of magnetic skyrmions in helimagnetic materials in response to spin-polarized currents.
\par
We consider a magnetic nanodisk of diameter \SI{120}{\nano\meter} ($x_1 x_2$-plane) and thickness $d=$ \SI{10}{\nano\meter} ($x_3$-direction).
We use the material parameters of iron-germanium (\ch{FeGe}), i.e., $\Ms=$ \SI{3.84e5}{\ampere\per\meter}, $A=$ \SI{8.78e-12}{\joule\per\meter}, $D=$ \SI{1.58e-3}{\joule\per\square\meter}, and $K=\SI{0}{\J\per\m^3}$; see, e.g., \cite{babcwcvhcsmf2017}.
The initial condition for our experiment is obtained by relaxing a uniform out-of-plane ferromagnetic state $\mm^0 \equiv (0,0,1)$ for \SI{2}{\nano\second}.
For the relaxation process, we choose the large value $\alpha=$ \num{1} for the Gilbert damping constant.
The resulting relaxed state is the skyrmion depicted in Figure~\ref{fig:helimag}.
\par
Starting from this configuration, we apply a perpendicular spin-polarized current pulse $\JJe(t) = (0,0,J(t))$ of maximum intensity $J_{\mathrm{max}}>0$ for \SI{150}{ps}; see Figure~\ref{fig:helimag_cur_avg}.
To model the resulting spin transfer torque, we include $\TT$ from~\eqref{eq:slonczewski}.
Then, we turn off the current density and let the system evolve for \SI{20}{\nano\second}.
In order to capture all possible excitation modes during the application of the pulse and the subsequent relaxation process, we set the value of the Gilbert damping constant to $\alpha=$ \num{0.002}, which is considerably smaller than the experimental value of $\alpha=$ \num{0.28} measured for \ch{FeGe}; see~\cite{babcwcvhcsmf2017}.
\par
In Figure~\ref{fig:helimag_cur_avg}, we plot the time evolution of the first component of the spatially averaged magnetization of the sample after a current pulse with $J_{\mathrm{max}}=$ \SI{1e12}{\ampere\per\meter\squared}, $\pp = (0,1,0)$, and $P =$ \num{0.4}.
The induced dynamics is a damped precession of the skyrmion around the center of the sample; see Figure~\ref{fig:helimag}.
We consider an unstructured tetrahedral mesh of the nanodisk generated by \ngsolve{}.
For a desired mesh size of $\SI{7.5}{\nm}$, the automatically generated mesh consists of $\num{35390}$ elements with maximum diameter $\meshsize=\SI{6.1}{\nm}$, $\num{8436}$ vertices, as well as $\num{9586}$ surface elements.
\begin{figure}[b]
\centering
\begin{minipage}{16cm}
\centering
\raisebox{-0.5\height}{\input{pics/helimag/mag_0.tex}}
\raisebox{-0.5\height}{\input{pics/helimag/mag_800.tex}}
\raisebox{-0.5\height}{\input{pics/helimag/mag_1600.tex}}
\raisebox{-0.5\height}{\input{pics/helimag/mag_2400.tex}}
\raisebox{-0.5\height}{\input{pics/helimag/mag_3200.tex}}
\raisebox{-0.5\height}{\input{pics/helimag/mag_4000.tex}}
\raisebox{-0.5\height}{\input{pics/helimag/mag_4800.tex}}
\raisebox{-0.5\height}{\input{pics/helimag/mag_5600.tex}}
\hspace*{-0.1cm}\raisebox{-0.5\height}{\input{pics/helimag/colorbar.tex}}
\end{minipage}
\caption{Snapshots of the skyrmion dynamics from Section~\ref{subsection:helimag}:
Magnetization in the $xy$-plane viewed from the top at different times.
Starting from the relaxed configuration at $t=\SI{0}{\ps}$, the skyrmion is deflected
from the center of the disk by a current pulse.
Then, the skyrmion oscillates around the center of the disk with an observed period
of approximately $\SI{400}{\ps}$.
Due to damping, over the relaxation period of $\SI{20}{\ns}$ the amplitude of the oscillations
decreases to almost zero, and the initial equilibrium configuration is restored.}
\label{fig:helimag}
\end{figure}
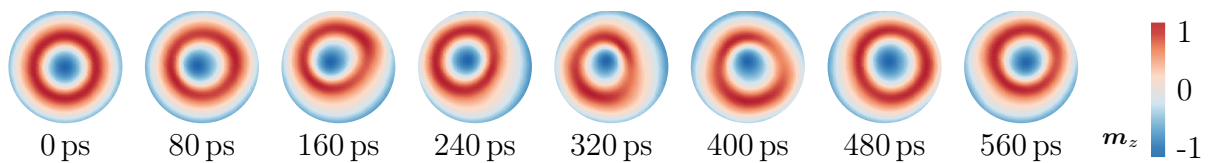
\begin{figure}[t]
\centering
\begin{minipage}{16cm}
\centering
\raisebox{-0.5\height}{\input{plots/helimag/pulse.tex}}
\hspace*{3mm}
\raisebox{-0.5\height}{\input{plots/helimag/avg_x.tex}}
\end{minipage}
\caption{Simulation of the skyrmion dynamics from Section~\ref{subsection:helimag}.
Structure of the applied current pulse (left).
Time evolution of the $x$-component of the spatially averaged magnetization (right).}
\label{fig:helimag_cur_avg}
\end{figure}
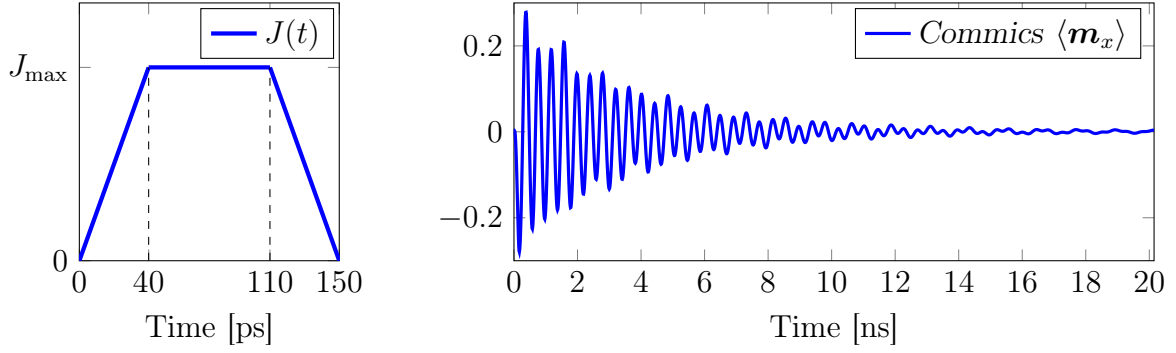
\begin{lstlisting}
from commics import *
# specify geometry and parameters
geo = Geometry(geometry=Disk(120.0, 10.0), meshSize=7.5, \
               forceNetgenMeshSize=True, scaling=1.0e-09)
par = Parameters(A=8.78e-12, Ms=3.84e+05, D=1.58e-03, dmCoupling="bulk", \
                 spintronicsCoupling="slonczewski", d=10.0e-09, P=0.4, \
                 p=(0.0, 1.0, 0.0), \
                 alpha=1.0, timeStepSize=1e-13, theta=1.0, m!0!=(0, 0, 1))
# define integrator, choose number of threads
helimag = Integrator(scheme=TPS!1!, geometry=geo, parameters=par, numthreads=8)
# choose to save magnetization as .vtk-file every X seconds, run simulation
helimag.RecordMagnetization(every=10.0e-12)
helimag.Integrate(duration=2.0e-9, relax=True)
# prepare JeMax, alpha, and points in time
helimag.SetParameter_alpha(0.002)
JeMax = 1.0e+12
T!0!, T!1!, T!2!, T!3! = 0, 40.0e-12, 110.0e-12, 150.0e-12
# set increasing current, run simulation
helimag.SetParameter_Je(lambda t,x,y,z : (t-T!0!)/(T!1!-T!0!)*JeMax)
helimag.Integrate(duration=T!1!-T!0!)
# set constant current, run simulation
helimag.SetParameter_Je(JeMax)
helimag.Integrate(duration=T!2!-T!1!)
# set decreasing current, run simulation
helimag.SetParameter_Je(lambda t,x,y,z : (T!3!-t)/(T!3!-T!2!)*JeMax)
helimag.Integrate(duration=T!3!-T!2!)
# final relaxation
helimag.Integrate(duration=20.0e-9, relax=True)
\end{lstlisting}
\clearpage
\appendix\section{Coupling of \ngsolve{} with \bempp{}}\label{appendix:fk}
In this section we show how \ngsolve{}, \bempp{}, and the \textsl{ngbem} module can be used to perform stray field computations following the approach proposed in~\cite{FK90}:
\begin{itemize}
  \item \ngsolve{} is used for mesh generation and the FEM problems.
  \item \bempp{} provides boundary integral operators and an interface for solving the BEM problem.
  \item \textsl{ngbem} provides the means to extract a boundary element mesh from the volume mesh together with a mapping between the corresponding degrees of freedom.
\end{itemize}
To test the procedure we consider the uniformly magnetized unit ball.
Then, the stray field is given by $\Hs(\MM) \equiv -\MM/3$ in $\Omega$.
The complete code is provided in the following Python script.
\begin{lstlisting}
# Demonstration of the Fredkin-Koehler approach for stray field computations
# using ngsolve, ngbem, bempp
import ngsolve, ngbem, bempp.api, commics, numpy
# mesh, femSpaces, GridFunctions
geo = commics.Geometry(commics.UnitBall(), meshSize=0.3)
geo.GetReady()
mesh = geo.GetMesh()
Vh!3! = ngsolve.VectorH!1!(mesh, order=1)
Vh = ngsolve.H!1!(mesh, order=1)
VhD = ngsolve.H!1!(mesh, order=1, dirichlet="bc_dirichlet")
m = ngsolve.GridFunction(Vh!3!)
m.components[0].vec.FV().NumPy()[:] = 1.0
u!1! = ngsolve.GridFunction(Vh)
u!2! = ngsolve.GridFunction(VhD)
hs = -u!1!.Deriv() - u!2!.Deriv()
# Neumann FEM problem
a!1! = ngsolve.BilinearForm(Vh)
a!1! += ngsolve.Laplace(1.0)
c!1! = ngsolve.Preconditioner(a!1!, "local")
a!1!.Assemble()
solverU!1! = ngsolve.CGSolver(mat=a!1!.mat, pre=c!1!.mat)
phi!1! = Vh.TestFunction()
f!1! = ngsolve.LinearForm(Vh)
f!1! += ngsolve.SymbolicLFI(phi!1!.Deriv() * m)
measOmega = ngsolve.Integrate(ngsolve.CoefficientFunction(1.0), mesh)
# BEM problem
bemSpace, op_NgToBem = ngbem.H!1!_trace(Vh)
op_NgToBem.eliminate_zeros()
bemToNgIdx = numpy.full(op_NgToBem.shape[0], fill_value = Vh.ndof, dtype=int)
bemToNgIdx[op_NgToBem.row] = op_NgToBem.col
bem_K = bempp.api.operators.boundary.laplace.double_layer(bemSpace, bemSpace, \
                                                          bemSpace)
bem_I = bempp.api.operators.boundary.sparse.identity(bemSpace, bemSpace, \
                                                     bemSpace)
bemRhsOp = bem_K - 0.5*bem_I
bemLhs = bem_I
# Dirichlet FEM problem
a!2! = ngsolve.BilinearForm(VhD)
a!2! += ngsolve.Laplace(1.0)
c!2! = ngsolve.Preconditioner(a!2!, "bddc")
a!2!.Assemble()
f!2! = ngsolve.LinearForm(VhD)
f!2!.Assemble()
bvp!2! = ngsolve.BVP(bf=a!2!, lf=f!2!, gf=u!2!, pre=c!2!)
# solve Neumann FEM problem
f!1!.Assemble()
u!1!.vec.data = solverU!1! * f!1!.vec
u!1!.vec.FV().NumPy()[:] -= ngsolve.Integrate(u!1!, mesh) / measOmega
# solve BEM problem
bem_u!1! = bempp.api.GridFunction(bemSpace, \
                                coefficients=u!1!.vec.FV().NumPy()[bemToNgIdx])
bem_rhs = bemRhsOp * bem_u!1!
bem_g = bempp.api.linalg.iterative_solvers.gmres(bemLhs, bem_rhs)[0]
# solve Dirichlet FEM problem
u!2!.vec.FV().NumPy()[bemToNgIdx] = bem_g.coefficients
bvp!2!.Do()
# check results
TOL = 0.01
HS = ngsolve.GridFunction(Vh!3!)
for d in range(3): HS.components[d].Set(hs[d])
passed = all(abs(-1/3 - HS.components[0].vec.FV().NumPy()) < TOL) \
         and all(abs(HS.components[1].vec.FV().NumPy()) < TOL) \
         and all(abs(HS.components[2].vec.FV().NumPy()) < TOL)
print("PASSED STRAYFIELD TEST:", passed)
\end{lstlisting}
\bibliographystyle{plain}
\bibliography{ref}
\end{document}

%% file: pythonListingStyle.tex
\newtoggle{InString}{}
\togglefalse{InString}
\newcommand*{\ColorIfNotInString}[1]{\iftoggle{InString}{#1}{\color{magenta}#1}}%
\newcommand*{\ProcessQuote}[1]{#1\iftoggle{InString}{\global\togglefalse{InString}}{\global\toggletrue{InString}}}
\lstdefinestyle{FormattedNumber}{%
    literate={'}{{{\ProcessQuote{'}}}}1
             {0}{{{\ColorIfNotInString{0}}}}{1}%
             {1}{{{\ColorIfNotInString{1}}}}{1}%
             {2}{{{\ColorIfNotInString{2}}}}{1}%
             {3}{{{\ColorIfNotInString{3}}}}{1}%
             {4}{{{\ColorIfNotInString{4}}}}{1}%
             {5}{{{\ColorIfNotInString{5}}}}{1}%
             {6}{{{\ColorIfNotInString{6}}}}{1}%
             {7}{{{\ColorIfNotInString{7}}}}{1}%
             {8}{{{\ColorIfNotInString{8}}}}{1}%
             {9}{{{\ColorIfNotInString{9}}}}{1}%
             {.0}{{{\ColorIfNotInString{.0}}}}{2}%
             {.1}{{{\ColorIfNotInString{.1}}}}{2}%
             {.2}{{{\ColorIfNotInString{.2}}}}{2}%
             {.3}{{{\ColorIfNotInString{.3}}}}{2}%
             {.4}{{{\ColorIfNotInString{.4}}}}{2}%
             {.5}{{{\ColorIfNotInString{.5}}}}{2}%
             {.6}{{{\ColorIfNotInString{.6}}}}{2}%
             {.7}{{{\ColorIfNotInString{.7}}}}{2}%
             {.8}{{{\ColorIfNotInString{.8}}}}{2}%
             {.9}{{{\ColorIfNotInString{.9}}}}{2}%
             {-0}{{{\ColorIfNotInString{-0}}}}{2}%
             {-1}{{{\ColorIfNotInString{-1}}}}{2}%
             {-2}{{{\ColorIfNotInString{-2}}}}{2}%
             {-3}{{{\ColorIfNotInString{-3}}}}{2}%
             {-4}{{{\ColorIfNotInString{-4}}}}{2}%
             {-5}{{{\ColorIfNotInString{-5}}}}{2}%
             {-6}{{{\ColorIfNotInString{-6}}}}{2}%
             {-7}{{{\ColorIfNotInString{-7}}}}{2}%
             {-8}{{{\ColorIfNotInString{-8}}}}{2}%
             {-9}{{{\ColorIfNotInString{-9}}}}{2}%
             {e+}{{{\ColorIfNotInString{e+}}}}{2}%
             {e-}{{{\ColorIfNotInString{e-}}}}{2}%
             ,
    escapeinside={!!},
}

%% file: pics/meshing/structured.tex
\begin{tikzpicture}[scale=3,tdplot_main_coords]
\coordinate (x0) at (0,0,0);
\coordinate (x1) at (1,0,0);
\coordinate (x2) at (0,1,0);
\coordinate (x3) at (0,0,1);
\coordinate (x4) at (1,1,0);
\coordinate (x5) at (1,0,1);
\coordinate (x6) at (0,1,1);
\coordinate (x7) at (1,1,1);
\coordinate (T1shift) at ($\shift*(x0)+\shift*(x1)$);
\coordinate (T2shift) at ($\shift*(x5)+\shift*(x1)$);
\coordinate (T3shift) at ($\shift*(x5)+\shift*(x7)$);
\coordinate (T4shift) at ($\shift*(x6)+\shift*(x7)$);
\coordinate (T5shift) at ($\shift*(x6)+\shift*(x2)$);
\coordinate (T6shift) at ($\shift*(x0)+\shift*(x2)$);
\draw[fill=yellow, fill opacity=.8] ($(x7)+(T4shift)$)--($(x6)+(T4shift)$)--($(x3)+(T4shift)$)--cycle;
\draw[fill=yellow, fill opacity=.5] ($(x7)+(T4shift)$)--($(x4)+(T4shift)$)--($(x3)+(T4shift)$)--cycle;
\draw[fill=yellow, fill opacity=.5] ($(x4)+(T4shift)$)--($(x6)+(T4shift)$)--($(x3)+(T4shift)$)--cycle;
\draw[dashed] ($(x4) + (T3shift)$)--($(x7) + (T3shift)$);
\draw[fill=blue!50!green, fill opacity=.8] ($(x7)+(T3shift)$)--($(x5)+(T3shift)$)--($(x3)+(T3shift)$)--cycle;
\draw[fill=blue!50!green, fill opacity=.5] ($(x4)+(T3shift)$)--($(x5)+(T3shift)$)--($(x3)+(T3shift)$)--cycle;
\draw[dashed] ($(x4) + (T2shift) $)--($(x5) + (T2shift) $);
\draw[fill=blue, fill opacity=.8] ($(x1)+(T2shift)$)--($(x5)+(T2shift)$)--($(x3)+(T2shift)$)--cycle;
\draw[fill=blue, fill opacity=.5] ($(x1)+(T2shift)$)--($(x4)+(T2shift)$)--($(x3)+(T2shift)$)--cycle;
\draw[dashed] ($(x4) + (T5shift) $)--($(x6) + (T5shift) $);
\draw[fill=red!50!yellow, fill opacity=.8] ($(x2)+(T5shift)$)--($(x6)+(T5shift)$)--($(x3)+(T5shift)$)--cycle;
\draw[fill=red!50!yellow, fill opacity=.5] ($(x2)+(T5shift)$)--($(x4)+(T5shift)$)--($(x3)+(T5shift)$)--cycle;
\draw[dashed] ($(x4) + (T6shift) $)--($(x2) + (T6shift) $);
\draw[fill=red, fill opacity=.8] ($ (x2) + (T6shift) $)--($(x0) + (T6shift) $)--($(x3) + (T6shift) $)--cycle;
\draw[fill=red, fill opacity=.5] ($(x4)+(T6shift)$)--($(x0)+(T6shift)$)--($(x3)+(T6shift)$)--cycle;
\foreach \i in {x0,x1,x3}
    \draw[dashed] ($(x4) + (T1shift) $)--($(\i) + (T1shift) $);
\draw[fill=red!50!blue, fill opacity=.8] ($ (x0) + (T1shift) $)--($(x1) + (T1shift) $)--($(x3) + (T1shift) $)--cycle;
\end{tikzpicture}

%% file: plots/sp4/avg_combined.tex
\begin{tikzpicture}
\pgfplotstableread{plots/sp4/avg-oommf.dat}{\oommfData}
\pgfplotstableread{plots/sp4/avg-structured_3nm_tps2ab.dat}{\tpsab}
\pgfplotstableread{plots/sp4/avg-unstructured_netgen3nm_mps.dat}{\mps}
\begin{axis}[
width = 160mm,
height = 80mm,
xlabel={Time [\si{\nano\second}]},
xmin=0,
xmax=3,
ymin=-1.1,
ymax=1.1,
legend columns=3,
legend style={/tikz/column 3/.style={column sep=5pt}},
]
\addplot[purple, very thick] table[x=t, y=mx]{\oommfData};
\addplot[blue, very thick,dashed] table[x=t, y=mx]{\tpsab};
\addplot[brown,very thick,dash pattern=on 3pt off 3pt on 1pt off 3pt] table[x=t, y=mx]{\mps};
\addplot[red, very thick] table[x=t, y=my]{\oommfData};
\addplot[cyan,dashed, very thick] table[x=t, y=my]{\tpsab};
\addplot[gray,very thick,dash pattern=on 3pt off 3pt on 1pt off 3pt] table[x=t, y=my]{\mps};
\addplot[magenta, very thick] table[x=t, y=mz]{\oommfData};
\addplot[teal,dashed, very thick] table[x=t, y=mz]{\tpsab};
\addplot[darkgray,very thick,dash pattern=on 3pt off 3pt on 1pt off 3pt] table[x=t, y=mz]{\mps};
\legend{
\oommf{} $\langle \mm_x \rangle$,
\texttt{TPS2AB} $\langle \mm_x \rangle$,
\texttt{MPS} $\langle \mm_x \rangle$,
\oommf{} $\langle \mm_y \rangle$,
\texttt{TPS2AB} $\langle \mm_y \rangle$,
\texttt{MPS} $\langle \mm_y \rangle$,
\oommf{} $\langle \mm_z \rangle$,
\texttt{TPS2AB} $\langle \mm_z \rangle$,
\texttt{MPS} $\langle \mm_z \rangle$,
}
\end{axis}
\end{tikzpicture}

%% file: pics/sp4/structured_3nm_tps2ab_xZeroColX.tex
   \begin{tikzpicture}[scale=0.9]
     \begin{axis}[
         tick align=outside, xtick pos=left, ytick pos=left,
         scale only axis, width=12cm, height=3cm,            
         ylabel style={yshift=-0.3cm},
         xmin=-250,xmax=250, ymin=-62.5,ymax=62.5,
         colormap name={mBlueToRed},
         colorbar style={
           at={($(parent axis.right of east)+(15,0)$)},
           anchor=west,
           xlabel=$\,\,\,\mm_x$,
           x label style={at={(axis description cs:0,0.5)},anchor=east},
           ytick={-1,0,1},
           yticklabels={{\bf -1},{\bf 0},{\bf 1}},
           width=0.3cm,
         },
         xtick = {-250, -150, -50, 50, 150, 250},
         xticklabels = {0, 100, 200, 300, 400, 500},
         x label style={at={(axis description cs:0.56,-0.2)},anchor=east},
         ytick = {-62.5, 62.5},
         yticklabels = {0, 125},
         y label style={at={(axis description cs:-0.065,0.7)},anchor=east},
         point meta min=-1, point meta max=1,
         view={0}{90},
         colorbar,
         xlabel=$x\,{[}\SI{}{\nm}{]}$,
         ylabel=$y\,{[}\SI{}{\nm}{]}$,
       ]
       \addplot3[surf, shader=interp] table{pics/sp4/structured_3nm_tps2ab_xZero_colorX.dat};
       \addplot[quiver={u=\thisrow{u},v=\thisrow{v}, scale arrows=12}, -stealth, thick] table{pics/sp4/structured_3nm_tps2ab_xZero_alignedArrows.dat};
     \end{axis}
   \end{tikzpicture}

%% file: pics/sp5/sp5_mag_0ps.tex
 \begin{tikzpicture}[scale=0.9]
     \begin{axis}[
         ticks=none,
         scale only axis, width=4cm, height=4cm,            
         xmin=-50,xmax=50, ymin=-50,ymax=50,
         colormap name={mBlueToRed},
         point meta min=-0.2, point meta max=1,
         xlabel={$\stackrel{\Large t\,=\,\SI{0.0}{\ns}}{\phantom{=}}$},
         view={0}{90},
       ]
       \addplot3[surf, shader=interp] table{pics/sp5/sp5_mag_0ps_color.dat};
       \addplot[quiver={u=\thisrow{u},v=\thisrow{v}, scale arrows=6}, -stealth, thick] table{pics/sp5/sp5_mag_0ps_gridArrows.dat};
     \end{axis}
   \end{tikzpicture}

%% file: pics/sp5/sp5_mag_500ps.tex
 \begin{tikzpicture}[scale=0.9]
     \begin{axis}[
         ticks=none,
         scale only axis, width=4cm, height=4cm,            
         xmin=-50,xmax=50, ymin=-50,ymax=50,
         colormap name={mBlueToRed},
         point meta min=-0.2, point meta max=1,
         xlabel={$\stackrel{\Large t\,=\,\SI{0.5}{\ns}}{\phantom{=}}$},
         view={0}{90},
       ]
       \addplot3[surf, shader=interp] table{pics/sp5/sp5_mag_500ps_color.dat};
       \addplot[quiver={u=\thisrow{u},v=\thisrow{v}, scale arrows=6}, -stealth, thick] table{pics/sp5/sp5_mag_500ps_gridArrows.dat};
     \end{axis}
   \end{tikzpicture}

%% file: pics/sp5/sp5_mag_1000ps.tex
 \begin{tikzpicture}[scale=0.9]
     \begin{axis}[
         ticks=none,
         scale only axis, width=4cm, height=4cm,            
         xmin=-50,xmax=50, ymin=-50,ymax=50,
         colormap name={mBlueToRed},
         point meta min=-0.2, point meta max=1,
         xlabel={$\stackrel{\Large t\,=\,\SI{1.0}{\ns}}{\phantom{=}}$},
         view={0}{90},
       ]
       \addplot3[surf, shader=interp] table{pics/sp5/sp5_mag_1000ps_color.dat};
       \addplot[quiver={u=\thisrow{u},v=\thisrow{v}, scale arrows=6}, -stealth, thick] table{pics/sp5/sp5_mag_1000ps_gridArrows.dat};
     \end{axis}
   \end{tikzpicture}

%% file: pics/sp5/sp5_mag_1500ps.tex
 \begin{tikzpicture}[scale=0.9]
     \begin{axis}[
         ticks=none,
         scale only axis, width=4cm, height=4cm,            
         xmin=-50,xmax=50, ymin=-50,ymax=50,
         colormap name={mBlueToRed},
         point meta min=-0.2, point meta max=1,
         xlabel={$\stackrel{\Large t\,=\,\SI{1.5}{\ns}}{\phantom{=}}$},
         view={0}{90},
       ]
       \addplot3[surf, shader=interp] table{pics/sp5/sp5_mag_1500ps_color.dat};
       \addplot[quiver={u=\thisrow{u},v=\thisrow{v}, scale arrows=6}, -stealth, thick] table{pics/sp5/sp5_mag_1500ps_gridArrows.dat};
     \end{axis}
   \end{tikzpicture}

%% file: pics/sp5/sp5_mag_2000ps.tex
 \begin{tikzpicture}[scale=0.9]
     \begin{axis}[
         ticks=none,
         scale only axis, width=4cm, height=4cm,            
         xmin=-50,xmax=50, ymin=-50,ymax=50,
         colormap name={mBlueToRed},
         point meta min=-0.2, point meta max=1,
         xlabel={$\stackrel{\Large t\,=\,\SI{2.0}{\ns}}{\phantom{=}}$},
         view={0}{90},
       ]
       \addplot3[surf, shader=interp] table{pics/sp5/sp5_mag_2000ps_color.dat};
       \addplot[quiver={u=\thisrow{u},v=\thisrow{v}, scale arrows=6}, -stealth, thick] table{pics/sp5/sp5_mag_2000ps_gridArrows.dat};
     \end{axis}
   \end{tikzpicture}

%% file: pics/sp5/sp5_mag_8000ps.tex
 \begin{tikzpicture}[scale=0.9]
     \begin{axis}[
         ticks=none,
         scale only axis, width=4cm, height=4cm,            
         xmin=-50,xmax=50, ymin=-50,ymax=50,
         colormap name={mBlueToRed},
         point meta min=-0.2, point meta max=1,
         xlabel={$\stackrel{\Large t\,=\,\SI{8}{\ns}}{\phantom{=}}$},
         view={0}{90},
       ]
       \addplot3[surf, shader=interp] table{pics/sp5/sp5_mag_8000ps_color.dat};
       \addplot[quiver={u=\thisrow{u},v=\thisrow{v}, scale arrows=6}, -stealth, thick] table{pics/sp5/sp5_mag_8000ps_gridArrows.dat};
     \end{axis}
   \end{tikzpicture}

%% file: pics/sp5/sp5_colorbar.tex
\begin{tikzpicture}
\begin{axis}[
    hide axis,
    scale only axis,
    height=0pt,
    width=0pt,
    colormap name={mBlueToRed},
    colorbar horizontal,
    point meta min=-0.2,
    point meta max=1,
    colorbar style={
        xlabel={\large $\quad\mm_z$},
        x label style={at={(axis description cs:1.6,0.05)},anchor=west},
        rotate=90,
        width=8.04cm,
        xtick={0,0.5,1},
        xticklabels={{\bf 0},{\bf 0.5},{\bf 1}},
        xticklabel style={
          anchor=west,
        },
    }]
    \addplot [draw=none] coordinates {(0,0)};
\end{axis}
\end{tikzpicture}

%% file: plots/sp5/averages.tex
\begin{tikzpicture}[scale=1]
\pgfplotstableread{plots/sp5/avg_oommf.dat}{\oommfData}
\pgfplotstableread{plots/sp5/avg_commics.dat}{\tpsab}
\begin{axis}[
width = 160mm,
height = 50mm,
xlabel={Time [\si{\nano\second}]},
xmin=0,
xmax=8,
ymax=0.4,
ymin=-0.4,
legend cell align={left},
legend style={/tikz/column 2/.style={column sep=5pt}},
legend pos=north east,
legend columns=2,
]
\addplot[purple,ultra thick] table[x=t, y=mx]{\oommfData};
\addplot[blue,dashed,ultra thick] table[x=t, y=mx]{\tpsab};
\addplot[red,ultra thick] table[x=t, y=my]{\oommfData};
\addplot[cyan,dashed,ultra thick] table[x=t, y=my]{\tpsab};
\legend{
\oommf{} $\langle \boldsymbol{m}_x \rangle$,
\ourmodule{} $\langle \boldsymbol{m}_x \rangle$,
\oommf{} $\langle \boldsymbol{m}_y \rangle$,
\ourmodule{} $\langle \boldsymbol{m}_y \rangle$,
}
\end{axis}
\end{tikzpicture}

%% file: plots/fmr/avg_y.tex
\begin{tikzpicture}
\pgfplotstableread{plots/fmr/avg_oommf.dat}{\oommfData}
\pgfplotstableread{plots/fmr/avg_y_commics_3.4nm.dat}{\tpsab}
\begin{axis}[
width = 160mm,
height = 50mm,
xlabel={Time [\si{\nano\second}]},
xmin=0,
xmax=2.5,
ymin=0.58,
ymax=0.594,
yticklabel style={/pgf/number format/.cd,fixed,precision=3},
]
\addplot[red,ultra thick] table[x=t, y=my]{\oommfData};
\addplot[cyan,dashed,ultra thick] table[x=t, y=my]{\tpsab};
\legend{
\oommf{} $\langle \mm_y \rangle$,
\ourmodule{} $\langle \mm_y \rangle$,
}
\end{axis}
\end{tikzpicture}

%% file: plots/fmr/dft_y.tex
\begin{tikzpicture}
\pgfplotstableread{plots/fmr/dft_y_oommf.dat}{\oommfData}
\pgfplotstableread{plots/fmr/dft_y_commics_3.4nm.dat}{\tpsab}

\begin{semilogyaxis}[
width = 150mm,
height = 50mm,
xlabel={Frequency [\si{\GHz}]},
ylabel={Density},
xmin=0,
xmax=20,
ymin=1e-5,
ymax=1,
]
\addplot[red,ultra thick] table[x=frequency, y=density]{\oommfData};
\addplot[cyan,dashed,ultra thick] table[x=frequency, y=density]{\tpsab};
\legend{
\oommf{} $S_y$,
\ourmodule{} $S_y$,
}
\end{semilogyaxis}
\end{tikzpicture}

%% file: pics/helimag/mag_0.tex
\begin{tikzpicture}[scale=1.0]
  \begin{axis}[
      ticks=none,
      axis line style={draw=none},
      scale only axis = true, width=1.5cm, height=1.5cm,            
      xmin=-60,xmax=60, ymin=-60,ymax=60,
      colormap name={mBlueToRed},
      view={0}{90},
      xlabel={$\SI{0}{\ps}$},
    ]
    \addplot3[surf, shader=interp] table{pics/helimag/mag_0_coarse_color.dat};
  \end{axis}
  \fill[white,even odd rule] (0.75cm,0.75cm) circle (0.75cm) (-0.015cm,-0.015cm) rectangle (1.515cm,1.515cm);
\end{tikzpicture}

%% file: pics/helimag/mag_800.tex
\begin{tikzpicture}[scale=1.0]
  \begin{axis}[
      ticks=none,
      axis line style={draw=none},
      scale only axis = true, width=1.5cm, height=1.5cm,            
      xmin=-60,xmax=60, ymin=-60,ymax=60,
      colormap name={mBlueToRed},
      view={0}{90},
      xlabel={$\SI{80}{\ps}$},
    ]
    \addplot3[surf, shader=interp] table{pics/helimag/mag_800_coarse_color.dat};
  \end{axis}
  \fill[white,even odd rule] (0.75cm,0.75cm) circle (0.75cm) (-0.015cm,-0.015cm) rectangle (1.515cm,1.515cm);
\end{tikzpicture}

%% file: pics/helimag/mag_1600.tex
\begin{tikzpicture}[scale=1.0]
  \begin{axis}[
      ticks=none,
      axis line style={draw=none},
      scale only axis = true, width=1.5cm, height=1.5cm,            
      xmin=-60,xmax=60, ymin=-60,ymax=60,
      colormap name={mBlueToRed},
      view={0}{90},
      xlabel={$\SI{160}{\ps}$},
    ]
    \addplot3[surf, shader=interp] table{pics/helimag/mag_1600_coarse_color.dat};
  \end{axis}
  \fill[white,even odd rule] (0.75cm,0.75cm) circle (0.75cm) (-0.015cm,-0.015cm) rectangle (1.515cm,1.515cm);
\end{tikzpicture}

%% file: pics/helimag/mag_2400.tex
\begin{tikzpicture}[scale=1.0]
  \begin{axis}[
      ticks=none,
      axis line style={draw=none},
      scale only axis = true, width=1.5cm, height=1.5cm,            
      xmin=-60,xmax=60, ymin=-60,ymax=60,
      colormap name={mBlueToRed},
      view={0}{90},
      xlabel={$\SI{240}{\ps}$},
    ]
    \addplot3[surf, shader=interp] table{pics/helimag/mag_2400_coarse_color.dat};
  \end{axis}
  \fill[white,even odd rule] (0.75cm,0.75cm) circle (0.75cm) (-0.015cm,-0.015cm) rectangle (1.515cm,1.515cm);
\end{tikzpicture}

%% file: pics/helimag/mag_3200.tex
\begin{tikzpicture}[scale=1.0]
  \begin{axis}[
      ticks=none,
      axis line style={draw=none},
      scale only axis = true, width=1.5cm, height=1.5cm,            
      xmin=-60,xmax=60, ymin=-60,ymax=60,
      colormap name={mBlueToRed},
      view={0}{90},
      xlabel={$\SI{320}{\ps}$},
    ]
    \addplot3[surf, shader=interp] table{pics/helimag/mag_3200_coarse_color.dat};
  \end{axis}
  \fill[white,even odd rule] (0.75cm,0.75cm) circle (0.75cm) (-0.015cm,-0.015cm) rectangle (1.515cm,1.515cm);
\end{tikzpicture}

%% file: pics/helimag/mag_4000.tex
\begin{tikzpicture}[scale=1.0]
  \begin{axis}[
      ticks=none,
      axis line style={draw=none},
      scale only axis = true, width=1.5cm, height=1.5cm,            
      xmin=-60,xmax=60, ymin=-60,ymax=60,
      colormap name={mBlueToRed},
      view={0}{90},
      xlabel={$\SI{400}{\ps}$},
    ]
    \addplot3[surf, shader=interp] table{pics/helimag/mag_4000_coarse_color.dat};
  \end{axis}
  \fill[white,even odd rule] (0.75cm,0.75cm) circle (0.75cm) (-0.015cm,-0.015cm) rectangle (1.515cm,1.515cm);
\end{tikzpicture}

%% file: pics/helimag/mag_4800.tex
\begin{tikzpicture}[scale=1.0]
  \begin{axis}[
      ticks=none,
      axis line style={draw=none},
      scale only axis = true, width=1.5cm, height=1.5cm,            
      xmin=-60,xmax=60, ymin=-60,ymax=60,
      colormap name={mBlueToRed},
      view={0}{90},
      xlabel={$\SI{480}{\ps}$},
    ]
    \addplot3[surf, shader=interp] table{pics/helimag/mag_4800_coarse_color.dat};
  \end{axis}
  \fill[white,even odd rule] (0.75cm,0.75cm) circle (0.75cm) (-0.015cm,-0.015cm) rectangle (1.515cm,1.515cm);
\end{tikzpicture}

%% file: pics/helimag/mag_5600.tex
\begin{tikzpicture}[scale=1.0]
  \begin{axis}[
      ticks=none,
      axis line style={draw=none},
      scale only axis = true, width=1.5cm, height=1.5cm,            
      xmin=-60,xmax=60, ymin=-60,ymax=60,
      colormap name={mBlueToRed},
      view={0}{90},
      xlabel={$\SI{560}{\ps}$},
    ]
    \addplot3[surf, shader=interp] table{pics/helimag/mag_5600_coarse_color.dat};
  \end{axis}
  \fill[white,even odd rule] (0.75cm,0.75cm) circle (0.75cm) (-0.015cm,-0.015cm) rectangle (1.515cm,1.515cm);
\end{tikzpicture}

%% file: pics/helimag/colorbar.tex
\begin{tikzpicture}[scale=1.0]

\begin{axis}[
    axis line style={white},
      view={0}{90},
    xmin=0,xmax=1, ymin=-1,ymax=1,
    point meta min=-1,
    point meta max=1,
    scale only axis = true,
    height=1.8cm,
    width=0.2cm,
    tick style = {color = black, draw opacity = 0},
    yticklabel pos=right,
    xmajorticks=false,
    ytick = {-0.85, 0, 0.85},
    yticklabels = {-1, 0, 1},
    xlabel = {\scriptsize $\mm_z$},
    xlabel style = {anchor = south east},
    xlabel style = {at={(axis description cs:0,0.0)}},
      ]
\addplot3[surf, shader=interp] {y}; 
\end{axis}

\end{tikzpicture}

%% file: plots/helimag/pulse.tex
\begin{tikzpicture}
\begin{axis}[
width = 50mm,
height = 50mm,
xlabel={Time [\si{\pico\second}]},
xmin=0,
xmax=150,
ymin=0,
ymax=80,
xtick={0,40,110,150},
xticklabels={0,40,110,150},
ytick={0,60},
yticklabels={0,$J_{\mathrm{max}}$},
]
\addplot[blue,ultra thick] coordinates {(0,0) (40,60)};
\addlegendentry{$J(t)$}
\addplot[blue,ultra thick] coordinates {(40,60) (110,60)};
\addplot[blue,ultra thick] coordinates {(110,60) (150,0)};
\addplot[dashed] coordinates {(40,0) (40,60)};
\addplot[dashed] coordinates {(110,0) (110,60)};
\end{axis}
\end{tikzpicture}

%% file: plots/helimag/avg_x.tex
\begin{tikzpicture}
\pgfplotstableread{plots/helimag/avg_x.dat}{\tps}
\begin{axis}[
width = 100mm,
height = 50mm,
xlabel={Time [\si{\nano\second}]},
xmin=0,
xmax=20.15,
ymin=-0.3,
ymax=0.3,
]
\addplot[blue,very thick] table[x=t, y=mx]{\tps};
\legend{
\ourmodule{} $\langle \mm_x \rangle$,
}
\end{axis}
\end{tikzpicture}